\documentclass[%
 reprint,
superscriptaddress,
 amsmath,amssymb,
 aps,
]{revtex4-2}

\usepackage{graphicx}
\usepackage{dcolumn}
\usepackage{bm}

\usepackage{physics}
\usepackage{upgreek}
\usepackage{gensymb}
\usepackage{units}
\usepackage{float}
\usepackage[colorlinks=true,linkcolor=gray,urlcolor=blue,citecolor=blue]{hyperref}

\usepackage{float}
\usepackage[commentmarkup=uwave,deletedmarkup=sout]{changes}
\definechangesauthor[name=Malte Grosche, color=blue]{fmg}
\begin{document}
\title{Magnetic signatures of pressure-induced multicomponent superconductivity in UTe$_2$}

\author{Zheyu Wu}
\author{Jiasheng Chen}
\author{Theodore. I. Weinberger}
\affiliation{Cavendish Laboratory, University of Cambridge, Cambridge, CB3 0HE, United Kingdom}

\author{Andrej Cabala}
\author{Vladim\'{i}r Sechovsk\'{y}}
\author{Michal Vali{\v{s}}ka}
\affiliation{Charles University, Faculty of Mathematics and Physics,\\ Department of
    Condensed Matter Physics, Ke Karlovu 5, Prague 2, 121 16, Czech Republic}

\author{Patricia L. Alireza}
\author{Alexander G. Eaton}
\email{age28@cam.ac.uk}
\author{F. Malte Grosche}
\email{fmg12@cam.ac.uk}
\affiliation{Cavendish Laboratory, University of Cambridge, Cambridge, CB3 0HE, United Kingdom}

\date{\today}

\begin{abstract}
    \noindent
    The phase diagram of the heavy fermion compound UTe$_2$ contains multiple superconducting phases, several of which show characteristics of odd-parity pairing. We have investigated the pressure dependence of the superconducting transition in high-quality crystals of UTe$_2$ by tracking its signature in the magnetic susceptibility $\chi(T)$. A single, sharp superconducting transition is observed at low pressures $<\unit[0.3]{GPa}$. At higher pressure, a second feature emerges in $\chi(T)$, which is located at the lower-temperature thermodynamic phase boundary previously identified in specific heat measurements. This second transition anomaly in $\chi(T)$ can be attributed to a step change in the London penetration depth, providing direct evidence for a change in the superconducting order parameter of UTe$_2$. Thermodynamic constraints suggest that the low temperature, high pressure superconducting state is distinct from zero pressure superconductivity as well as from the high pressure, high temperature superconducting state, raising the possibility of multicomponent superconductivity in high pressure UTe$_2$.
\end{abstract}

\maketitle
\noindent
Few materials possess phase diagrams containing several distinct superconducting states.
Examples include the heavy fermion metals UPt$_3$ \cite{FisherUPt3PhysRevLett.62.1411,HasselbackUPt3PhysRevLett.63.93,TailleferUPt3RevModPhys.74.235} and U$_{1-x}$Th$_x$Be$_{13}$ \cite{Ott85PhysRevB.31.1651,Kim91PhysRevB.44.6921}, both of which exhibit two anomalies in their specific heat as a function of temperature, $C(T)$, signalling a second-order thermodynamic phase transition between distinct superconducting states. In the ferromagnet URhGe, superconductivity is first suppressed and then re-emerges in applied field along the hard axis \cite{levy2005URhGe,yelland2011URhGe,Sherkunov18PhysRevLett.121.097001}, and in CeRh$_2$As$_2$, a second superconducting state with upper critical field $B_{c2}$ far exceeding the conventional Pauli limit is likewise induced in fields along the hard axis \cite{khim2021scienceCeRh2As2,brandoQUADPhysRevX.12.011023,KonstantinPhysRevB.107.L220504,ChajewskiPhysRevLett.132.076504}.

The heavy fermion paramagnet UTe$_2$ also changes into a different superconducting state for a magnetic field, $B$, applied along the hard magnetic direction, which in UTe$_2$ is the crystalline $b$-axis \cite{Ran2019Science,Ranfieldboostednatphys2019}. Heat capacity $C(T,B)$ \cite{Rosuel23} and NMR measurements \cite{Kinjo23PRB} have confirmed that this field-induced superconducting state SC2 is a distinct thermodynamic phase from the zero-field superconducting ground state SC1. Nuclear magnetic resonance (NMR) studies in magnetic fields $\parallel b$  showed no significant change in the Knight shift -- and hence in the local spin susceptibility --  on crossing into SC2 from the normal state. A change in the Knight shift was, however, observed, when UTe$_2$ changes from SC2 to SC1 \cite{Kinjo23PRB}.
This indicates that the dominant spin component of the spin-triplet Cooper pair rotates from being $\parallel a$ in SC1 to $\parallel b$ in SC2 \cite{Kinjo23PRB}. 
At high fields, SC2 is abruptly truncated at a ceiling field $B_m \approx$~35~T, which marks the metamagnetic transition to a field-polarised (FP) paramagnetic phase \cite{Miyake2019,LinNevidomskyyPaglione20,Aoki_UTe2review2022}. The FP state in turn hosts yet another superconducting phase (SC3) for specific orientations of the magnetic field, which persists up to $B \approx$~70~T \cite{Ranfieldboostednatphys2019,helm2024,qcl}.

\begin{figure}[t]
    \centerline{\includegraphics[width=0.85\columnwidth]{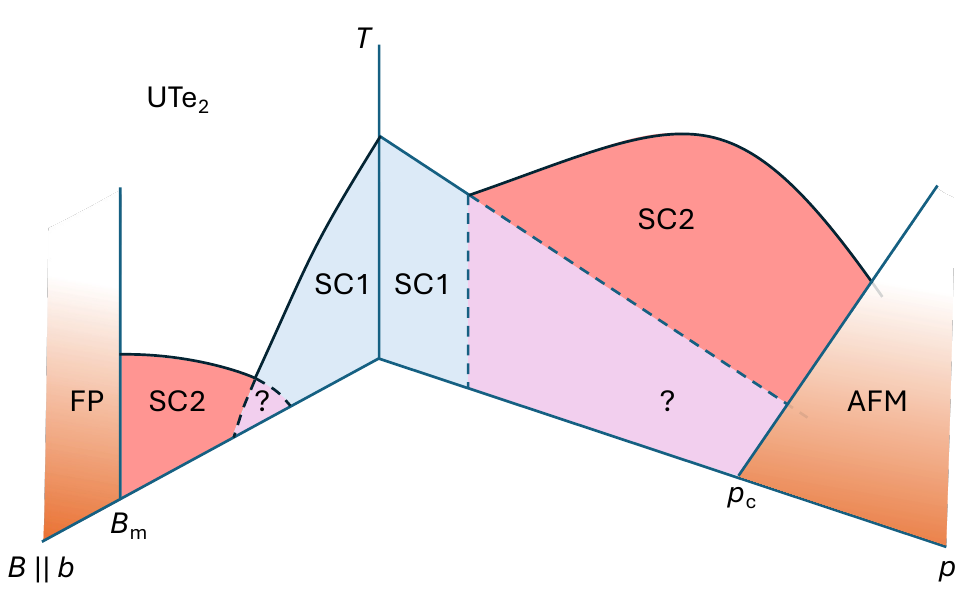}}
    \caption{Schematic pressure $p$, field $B$ ($\parallel b$), temperature $T$ phase diagram in UTe$_2$, illustrating that both pressure and field $\parallel b$ first decrease, then increase $T_c$, until superconductivity is cut off as magnetic order sets in beyond either a threshold pressure $p_c$ (antiferromagnetism AFM) or threshold field $B_m$ (field-polarised state FP). The dashed lines indicate further transitions within the superconducting state. The nature of the phases marked `?' is unclear.}
    \label{fig:SchematicDia}
\end{figure}
Some first-generation UTe$_2$ samples displayed two heat capacity anomalies at ambient pressure and in zero magnetic field, prompting the proposal that SC1 possesses a multicomponent chiral order parameter \cite{Hayes-junk-science,KapitulnikPhysRevB.105.024521}. However, subsequent measurements on next-generation, higher quality samples instead exhibited a single sharp transition in $C(T)$ at ambient pressure \cite{PhysRevMaterials.6.073401MSF_UTe2,Aoki_UTe2review2022,ajeesh2023fate,Rosa2022}. A multicomponent order parameter in SC1 is expected to cause discontinuities in the elastic shear moduli. Absence of such discontinuities at $T_c$ in both single- and double-transition samples probed by ultrasound argues strongly against the scenario of a multicomponent order parameter for the SC1 phase \cite{theuss24}.

\begin{figure*}[t!]
    \includegraphics[width=1\linewidth]{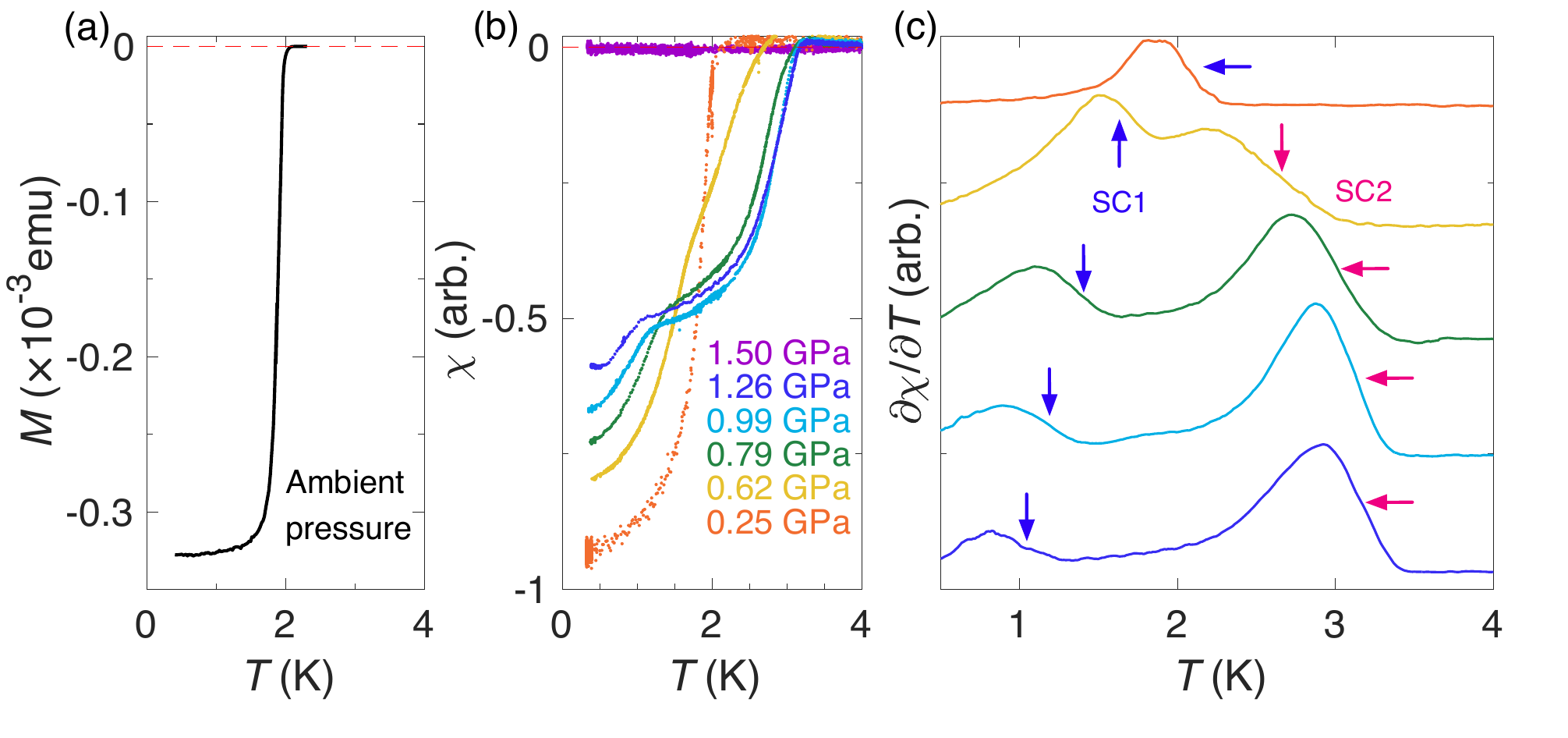}
    \caption{(a) dc magnetic moment, $M$, of a UTe$_2$ single crystal at ambient pressure plotted as a function of temperature, $T$. The sample was mounted on a quartz holder and measured in a magnetic field of 1.0 mT. (b) ac magnetic susceptibility, $\chi$, of a UTe$_2$ specimen in a MAC measured at incremental pressure points as indicated. Arrows mark superconducting transition anomalies at $T_{c1}$ (blue arrow) and $T_{c2}$ (magenta arrow). (c) Temperature derivatives of the data in panel b. A singular minimum is observed in the 0.25~GPa curve, identified with an arrow marking the SC1 transition. All higher pressure points exhibit an additional minimum, labelled as the SC2 transition.}
    \label{fig:chi-curves}
\end{figure*}

With increasing pressure $p$, the superconducting transition temperature $T_c$ decreases initially and then rises again, peaks at about \unit[1.2]{GPa} and vanishes inside the pressure-induced magnetically orderd phase. Magnetic transition signatures are detected above about \unit[1.5]{GPa} but they extrapolate to zero temperature at a critical pressure $p_c \simeq \unit[1.3]{GPa}$, which roughly coincides with the peak $T_c$. Heat capacity measurements under pressure \cite{Braithwaite2019,Thomas2020,Aoki_UTe2review2022} detect a second, lower transition anomaly, which continues the initial decrease of $T_c$ with pressure and extrapolates to zero at high pressure (\autoref{fig:SchematicDia}). An isolated high pressure NMR measurement at 12 kbar \cite{kinjo2023SciAdv} indicates no Knight shift change at the upper transition, whereas the Knight shift decreases below the lower transition, suggesting two distinct superconducting states. Furthermore, the metamagnetic ceiling field $B_m$ decreases with pressure.

Here, we investigate the high pressure phase diagram in UTe$_2$ using ac magnetic susceptibility $\chi(T)$ measurements in ultra-clean UTe$_2$ single crystals. We find that (i) the lower transition anomaly induced at finite pressure, which was previously only observed in $C(T)$, is associated with a pronounced decrease in $\chi(T)$, (ii) this decrease can be attributed to a reduction in the superconducting penetration depth $\lambda$, demonstrating that (iii) the lower transition marks a change in the superconducting state \cite{prozorov2006magnetic,Mueller_PhysRevB.108.144501}. Our findings confirm that SC1 and SC2 identified in applied field at ambient pressure connect to the low and high temperature superconducting states in pressurised UTe$_2$. They invite a more detailed study of the SC2 state and the high field polarised state under pressure, where the field ranges required to reach these states are significantly decreased.




\textit{Methods} -- High quality UTe$_2$ single crystals were grown by the molten salt flux (MSF) technique \cite{PhysRevMaterials.6.073401MSF_UTe2} in excess uranium.
All samples were selected from the same growth batch from which we previously drew crystals for quantum oscillation measurements \cite{eaton2024}, demonstrating the high sample quality. $\chi(T)$ measurements were performed using a microcoil setup in moissanite anvil cells (MACs) \cite{RevSciIns_Pressure_ACMS} with glycerol as the pressure medium.
In this setup, a platelet sample with approximate thickness $<\unit[50]{\upmu m}$ and diameter $\approx\unit[100]{\upmu m}$ was loaded in a 10-turn microcoil with axis aligned along the crystallographic $c$-direction. An ICEOxford $^3$He dipper probe was used to access temperatures down to $\unit[0.35]{K}$. $\chi(T)$ was inferred from the quadrature component of a lock-in measurement of the voltage induced across the microcoil, with the excitation current of 1~mA at 1.33~kHz applied to a concentric 170-turn outer drive coil. DC magnetic moment measurements at ambient pressure were performed in a Quantum Design Magnetic Property Measurement System using a $^3$He module. Heat capacity $C(T)$ was measured by the ac-calorimetry method in a piston cylinder cell (PCC) using Daphne oil 7474 as pressure medium \cite{klotz2009daphne}. Pressure was determined by ruby fluorescence \cite{ruby_pressure} in anvil cell measurements and from the resistive $T_c$ of Pb \cite{lead_pressure} in piston cylinder cell measurements.

\textit{Results} -- Figure~\ref{fig:chi-curves} shows $\chi(T)$ of compressed UTe$_2$ at various hydrostatic pressure values as indicated. At ambient pressure (black curve, \autoref{fig:chi-curves}a) and 0.25~GPa (orange points, \autoref{fig:chi-curves}b) a single transition is observed, which is identified by a sharp sudden decrease of the susceptibility. This is as expected for the diamagnetic screening of magnetic flux by supercurrents upon cooling below the superconducting critical temperature, $T_{\text{c}}$. By contrast, for the $\chi(T)$ curves at $p >$~0.25~GPa, \textit{two} such transition anomalies are observed, as clearly identified in $\nicefrac{\partial \chi}{\partial T}$, labelled as $T_\text{c1}$ and $T_\text{c2}$ in \autoref{fig:chi-curves}c.

\begin{figure}[t!]
    \includegraphics[width=1\linewidth]{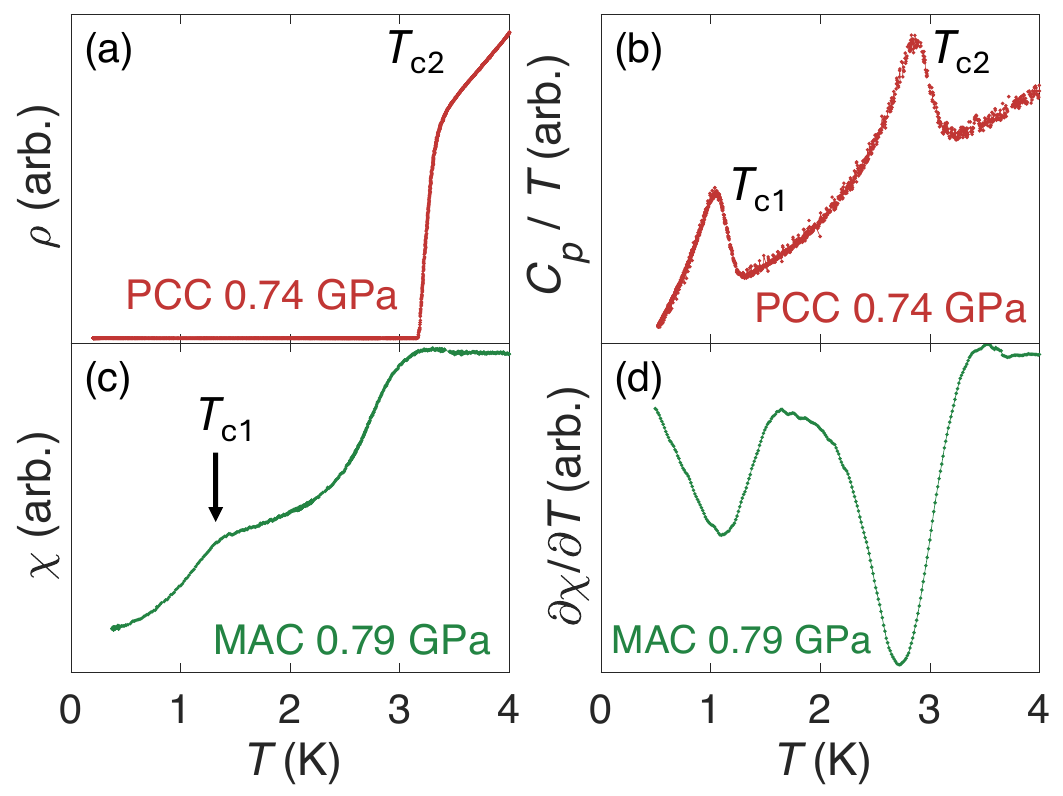}
    \caption{Resistivity $\rho$ (a) and specific heat divided by temperature $C_p/T$ (b) measured in a piston cylinder cell (PCC) at $p =$~0.74~GPa compared with the susceptibility $\chi$ (c) and $\partial{\chi}/\partial{T}$ (d) for a sample in a moissanite anvil cell (MAC) at $p =$~0.79~GPa. The two anomalies in $\chi$ clearly correspond to the two anomalies in $C_p$.}
    \label{fig:comparison}
\end{figure}

The presence of two distinct superconducting states in compressed UTe$_2$ for $p \gtrsim \unit[0.3]{GPa}$ has previously been reported from measurements of $C(T)$ on samples grown by chemical vapour transport (CVT) \cite{Braithwaite2019,Thomas2020,Aoki_UTe2review2022}. To determine whether the two features in $\chi(T)$ at $p > \unit[0.25]{GPa}$ correspond to bulk thermodynamic superconducting transitions, we also performed measurements of the resistivity, $\rho (T)$, and $C(T)$ on an MSF UTe$_2$ specimen.
Figure~\ref{fig:comparison} compares $\chi$ and $\nicefrac{\partial \chi}{\partial T}$ for a sample measured in a moissanite anvil cell at $p=\unit[0.79]{GPa}$ with $\rho (T)$ and $C(T)$ measured on a different sample in a piston-cylinder cell at 0.74~GPa. The resistivity falls to zero at the onset of the first anomaly in $C(T)$ at $T \approx$~3.16~K and remains zero below this temperature. The ac magnetic susceptibility, by contrast, shows two consecutive drops, and the two features in $\nicefrac{\partial\chi}{\partial T}$ clearly correspond to the two $C(T)$ anomalies. This agreement between heat capacity and magnetic susceptibility anomalies in our MSF-grown UTe$_2$ strongly suggests that the lower-temperature $\chi(T)$ anomaly reflects the transition previously identified solely in high-pressure heat capacity measurements in CVT-grown UTe$_2$ \cite{Braithwaite2019,Thomas2020,Aoki_UTe2review2022}.

\begin{figure}[t]
    \includegraphics[width=1\linewidth]{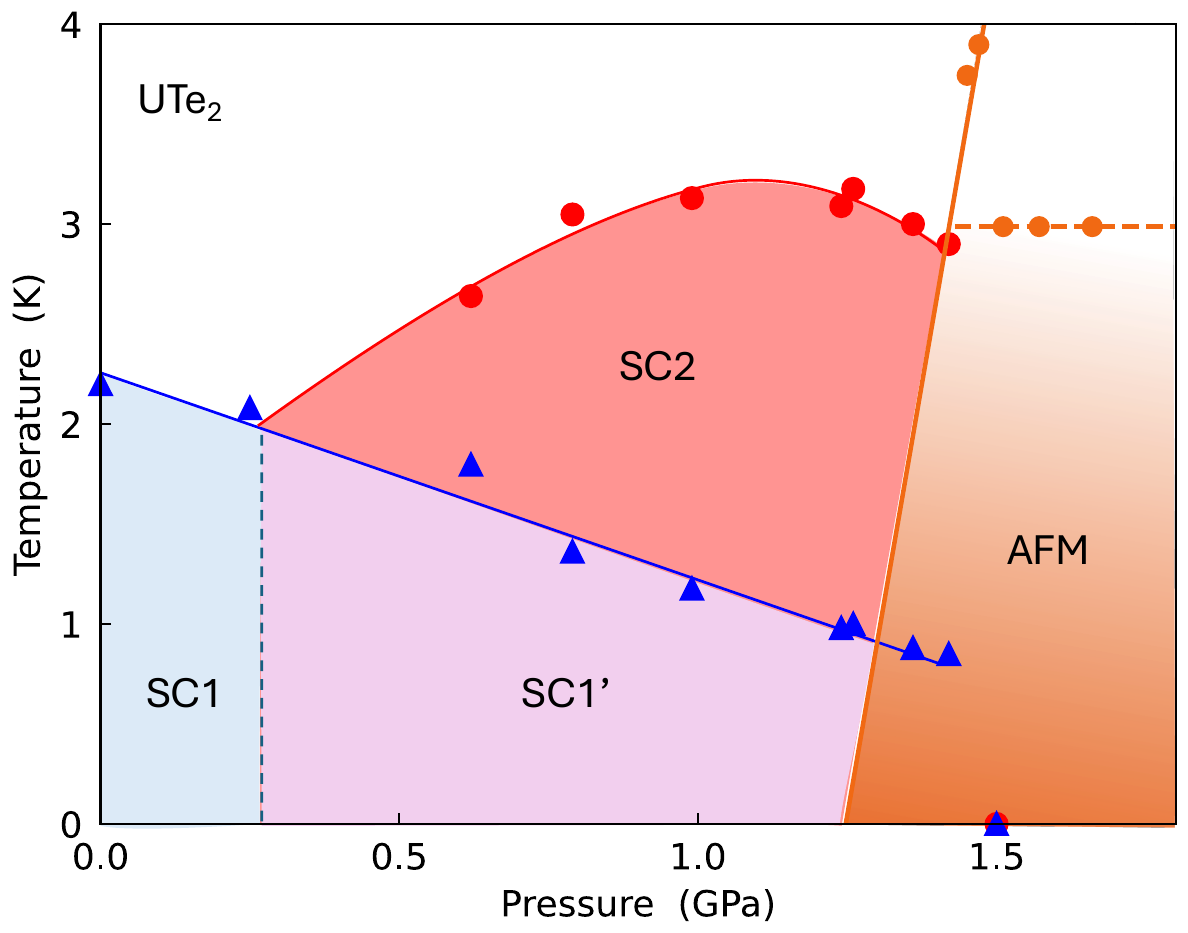}

    \caption{Temperature--pressure phase diagram of UTe$_2$. SC1 (blue triangles) and SC2 (red circles) transition temperatures are from our ac magnetic susceptibility measurements; the orange circles marking the antiferromagnetic (AFM) state are reproduced from ref. \cite{Thomas2020}, with AFM transition lines representing linear fits to the data. The dashed line at $p$~=~0.29~GPa is located where ref. \cite{Thomas104.224501} identifies the first signature of the SC2 phase. 
    }
    \label{fig:phasediag}
\end{figure}

\autoref{fig:phasediag} summarises the pressure dependence of the $\chi(T)$ transition anomalies in UTe$_2$ in a $T$--$p$ phase diagram for $B =$~0~T. Coexistence of SC1 \& SC2 is indicated above $\unit[0.3]{GPa}$, similar to \cite{kinjo2023SciAdv}.
We find that the higher $T_c$ value at ambient pressure of our MSF-grown UTe$_2$ compared to that of CVT-grown UTe$_2$ translates to systematically higher $T_c$ values for both the SC1 and SC2 states at all pressures up to the critical pressure $p_c \approx$ 1.46~GPa, beyond which superconductivity disappears abruptly (see comparison to literature values in Supplementary Materials).

Superconductivity disappears abruptly as the pressure is tuned through $p_c$, in line with prior studies on CVT UTe$_2$ samples \cite{Braithwaite2019,Thomas2020,Aoki_UTe2review2022}. No signatures of superconductivity are observed at 1.50 GPa, whereas two transitions can still be resolved at 1.42 GPa ($T_{\text{c2}} =$~2.92~K and $T_{\text{c1}} =$~0.85~K). At pressures exceeding \unit[1.42]{GPa},  $\chi(T)$ is flat and featureless below 4~K (\autoref{fig:chi-curves}b).
The sudden disappearance of superconducting transition anomalies and lack of other (i.e. magnetic) transition anomalies in the uniform susceptibility is consistent with antiferromagnetic order above $p_c$, as recently identified by neutron scattering measurements \cite{knafo2023incommensurate}.

\textit{Discussion} -- Observing a second drop in $\chi(T)$ at the lower transition temperature $T_{c1}$ in addition to the drop at the upper transition at $T_{c2}$ is surprising, because the onset of perfect conductivity at $T_{c2}>T_{c1}$ might be expected to lead to complete screening of the interior of the sample, preventing further changes in the susceptibility on cooling. This expectation falls short, however, because the sample size is small enough (radius $R \simeq \unit[50]{\upmu m}$, thickness $t < R$) for the penetration depth $\lambda$ to matter. Magnetic fields penetrate the perimeter of the sample within a surface layer of thickness $\simeq \lambda$, leaving a screened volume $\simeq \pi(R-\lambda)^2 t$. Moreover, our thin platelet crystal is probed by a coil wound around its perimeter, causing significant demagnetizing fields. Writing the signal as $\propto \text{screened volume }\times \left(-1/(1-N)\right)$, the demagnetizing factor $N$ can be approximated as $N^{-1} \simeq 1+0.8 t/(R-\lambda)$ \cite{prozorov18a}. The detected signal then scales not with the excluded volume ($\pi  t (R-\lambda)^2$) but with $(R-\lambda)^3$, so the relative signal change is $3 \lambda /R$ to first order in $\lambda/R$. Because $\lambda$ diverges at $T_c$ and is of order 1~$\upmu$m at low $T$ in UTe$_2$ (e.g. \cite{Metz_therm-trans2019}), the susceptibility signal can be expected to reflect changes in $\lambda$ in these measurements.

A similar double anomaly in $\chi(T)$ has indeed been observed in UPt$_3$ at ambient pressure, where the location of the kink in the UPt$_3$ phase diagram was found to coincide with the bulk thermodynamic phase transition between the superconducting A and B phases. This was interpreted in terms of a change in $\lambda$ at the transition \cite{TailleferUPt3RevModPhys.74.235,BrulsPhysRevLett.65.2294,SignorePhysRevB.52.4446,SCHOTTL2000}. A similar result has also been reported for U$_{0.97}$Th$_{0.03}$Be$_{13}$ \cite{ShimizuPRB96.100505}. Motivated by these reports, we interpret the lower-$T$ kink anomaly in $\chi(T)$ as indicating a further reduction in the London penetration depth below the value reached within SC2 at this temperature. This result indicates a transition in the superconducting order parameter of UTe$_2$ on crossing from SC2 to SC1. Our findings and their interpretation suggest that a more detailed study of the penetration depth in pressurised UTe$_2$ to determine absolute values for $\lambda(T)$ could provide important information about the nature of the two superconducting states SC1 and SC2.







The symmetry of the superconducting order parameter in the SC1 state of UTe$_2$ remains an open and actively pursued question. Ultrasound measurements \cite{theuss24} suggest that the SC1 order parameter is single-component in character, but there is conflicting evidence regarding the existence and location of gap nodes:
Measurements in CVT samples of thermal conductivity \cite{Metz_therm-trans2019}, NMR \cite{Fujibayashi_JPSP2022}, specific heat \cite{Kittaka_PhysRevResearch.2.032014}, scanning SQUID \cite{Molersuperfluid} and pulse-echo ultrasound \cite{theuss24}, as well as recent thermal transport measurements on a new sample generation grown in Te-flux were all interpreted in terms of point-nodes located in the $k_c = 0$ plane \footnote{The authors of ref. \cite{Molersuperfluid} also noted their data would be consistent with a highly anisotropic full gap.}, which is consistent with the Fermi surface geometry of quasi-2D pockets that run along the $c$ direction \cite{AokidHvA_UTe2-2022,theo2024,eaton2024,weinberger2024pressureenhanced,hayes2024robust}.
By contrast, subsequent NMR \cite{matsumura2023NMR-aoki} and thermal conductivity \cite{suetsugu2024fully} studies in high-quality MSF samples have instead been taken to indicate a fully gapped pairing state. The question of whether the gap function contains point nodes -- and if so, where on the Fermi surface they are located -- thus remains the topic of active debate. 

For the SC2 state, theoretical studies have likewise proposed a number of possible order parameter symmetries, including nodal \cite{LinNevidomskyyPaglione20} and fully gapped \cite{Ishizuka_PhysRevLett.123.217001} odd-parity states. Another study \cite{IshizukaPRB.103.094504} suggested that the superconducting state accessed for $B \parallel b$ at ambient pressure is separate to that found under pressure in ambient magnetic field, and that the pressurized state may be even-parity in character. However, recent NMR studies under applied magnetic field \cite{tokunaga2023longitudinal,Kinjo23PRB} and hydrostatic pressure \cite{kinjo2023SciAdv} have reported strong similarities in the NMR spectra of these two phases, including a negligible change in the Knight shift at $T_\text{c2}$, preferential spin alignment along the $b$-axis, and increasing $T_\text{c2}$ with increasing $b$-axis local spin susceptibility  under either pressure or magnetic field.
Each of these observations strongly suggest that the high-field and high-pressure superconducting phases both labelled SC2 in \autoref{fig:SchematicDia} indeed host the same odd-parity superconducting state.

Our observation of a transition in the superconducting order parameter of UTe$_2$, manifested by an anomaly in the London penetration depth at the SC1--SC2 transition, raises the question about the nature of the superconducting state in the region labelled SC1' in \autoref{fig:phasediag}. Thermodynamic arguments require a further transition line (dashed line in \autoref{fig:phasediag}) separating SC1' from SC1 \cite{yip91}, causing SC1' to be distinct from both SC1 and SC2. This favours the emergence of multicomponent superconductivity in SC1' out of a combination of the order parameters in SC1 and SC2. High pressure ultrasound studies and quantitative measurements of the London penetration depth at high pressure \cite{giannetta2022london}, will be of great value in resolving the nature of the superconducting order parameter in SC1'.

\vspace{-0mm}
\begin{acknowledgments} 
    We are grateful to A. Carrington, M.J. Grant, D.V. Chichinadze, D. Shaffer, T. Hazra and A.J. Hickey for stimulating discussions. This project was supported by the EPSRC of the UK (grants EP/X011992/1 \& EP/R513180/1). Crystal growth and characterization were performed in MGML (mgml.eu), which is supported within the program of Czech Research Infrastructures (project no. LM2023065). We acknowledge financial support by the Czech Science Foundation (GACR), project No. 22-22322S. A.G.E. acknowledges support from the Henry Royce Institute for Advanced Materials through the Equipment Access Scheme enabling access to the Advanced Materials Characterisation Suite at Cambridge, grants EP/P024947/1, EP/M000524/1 \& EP/R00661X/1; and from Sidney Sussex College (University of Cambridge).
\end{acknowledgments}

\bibliography{UTe2}
\end{document}